# EthicAlly: a Prototype for AI-Powered Research Ethics Support for the Social Sciences and Humanities

Author:

Dr Steph Grohmann

Ethics & Diversity Hub, Ludwig Boltzmann Gesellschaft
Nussdorferstrasse 64, Vienna, Austria

steph.grohmann@lbg.ac.at

Abstract

In biomedical science, review by a Research Ethics Committee (REC) is an indispensable way of protecting human subjects from harm. However, in social science and the humanities, mandatory ethics compliance has long been met with scepticism as biomedical models of ethics can map poorly onto methodologies involving complex socio-political and cultural considerations. As a result, tailored ethics training and support as well as access to RECs with the necessary expertise is lacking in some areas, including parts of Europe and low- and middle-income countries. This paper suggests that Generative AI can meaningfully contribute to closing these gaps, illustrating this claim by presenting EthicAlly, a proof-of-concept prototype for an AI-powered ethics support system for social science and humanities researchers. Drawing on constitutional AI technology and a collaborative prompt development methodology, EthicAlly provides structured ethics assessment that incorporates both universal ethics principles and contextual and interpretive considerations relevant to most social science research. In supporting researchers in ethical research design and preparation for REC submission, this kind of system can also contribute to easing the burden on institutional RECs, without attempting to automate or replace human ethical oversight.

## 1. Introduction

In biomedical science, institutional ethical oversight by Institutional Review Boards (IRBs) or Research Ethics Committees (RECs) is an indispensable component of ethical research with human participants. Due to the grave consequences of historical human rights violations in experimental research, ethics review for biomedical studies is today legally mandated in most jurisdictions. However, in social science and applied humanities research, including psychology and non-clinical research in the health sciences, the situation is more heterogeneous. This paper refers to these approaches collectively as "social science", on the basis that they most often use social scientific methods such as surveys, interviews, focus groups, and ethnographic methods.

The introduction of comprehensive REC oversight across all scientific fields, including those using social science methods, has been met with scepticism since its inception (Hammersley 2009; Schrag 2011). Critics have pointed out that ethics frameworks and review processes designed for biomedical science map poorly onto the relatively more heterogenous and flexible research designs used by social scientists, that institutional RECs are often insufficiently prepared to evaluate qualitative research designs, that mandatory REC oversight reduces ethical thinking to a bureaucratic "tick box exercise", and other real or perceived flaws. Despite these critiques, however, institutional mandates as well as demands from journal publishers and funders mean that today ethics planning in research design and preparation for ethics review are an inescapable reality for social scientists (Wassenaar&Marmotte 2012). In some areas, including parts of Europe but also in low- and middle-income countries worldwide, this development has created a mismatch between available infrastructure for ethics training, support and formal review for social science, and the demands of globally leading research and publication. These disparities threaten to disadvantage the career progression of social scientists compared to biomedical researchers, as well as exacerbating existing research inequities between regions (Sina&Tierney 2024).

This paper suggests that the use of Generative AI (GenAI) could make a meaningful contribution to addressing some of these disparities. To illustrate this point, the author introduces EthicAlly, a proof-of-concept prototype for an AI-powered research ethics support system for the social sciences and humanities. EthicAlly was developed at the Ethics & Diversity Hub of the Ludwig Boltzmann Gesellschaft, a publicly funded research institution in Austria, by the author, who is also head of the institution's REC. The system is intended to aid professional and student researchers in designing research that meets general and discipline-specific ethical standards, and preparing for REC submission by flagging issues a committee would likely identify. Its primary aim is to help avoid harm to research participants through raising researcher awareness of potential ethical issues, while also enabling researchers to navigate the ethics review process better and faster.

Similar systems have been developed for clinical research (Godwin et.al 2024), highlighting the contribution of REC preparation workload to researcher burnout and demonstrating measurable positive impacts of AI support in REC submission planning. Moreover, while preparation for clinical RECs has different requirements than for social scientific approaches, these interventions also demonstrate the potential of AI to indirectly support the work of REC members by reducing workload and speeding up review turnaround, without making any intervention in the review process or REC decision-making itself.

The author explicitly does not make any claims about using AI to partially or completely automate REC oversight, as suggested by some recent proposals (Porsdam Mann et.al. 2024). While this may be technically feasible, there are also substantial ethical problems with this approach as discussed below. The purpose of EthicAlly is strictly confined to enabling researchers to do more of the pre-submission work of addressing ethical issues, and in this way also ease the burden on RECs while maintaining human expertise as the basis for ultimate ethical decision-making.

## 2. Background

The introduction of REC oversight in social science fields has long been controversial. While few would argue that the ethical principles enshrined in modern ethics codices are unimportant, their application in fields working with social scientific methods is often less than straightforward. From the 1960s onwards, researchers have voiced concerns about an overly rigid application of ethical principles such as Informed Consent, a lack of REC expertise in specific fields of research, the potential for human bias to distort review processes, and the risk of bureaucratic compliance replacing genuine ethical debate (Schrag 2011). Critics have further pointed out that ethical frameworks and review processes designed for biomedical science do not always map well onto social scientific approaches, especially where qualitative methodologies are concerned (Moriña 2021; Taquette 2022; Dingwall 2008; 2012), or where research concerns sensitive topics (Winter & Gundur 2024). As a result, the expansion of REC oversight in these areas of study has been called a form of "ethics creep" (Haggerty 2004; White 2007) and is to this day met with persistent scepticism or outright rejection by some social science researchers who argue that biomedical ethics frameworks are poorly suited to their methodological approaches and research contexts (Carniel et al 2023). Especially fields like anthropology, the author's doctoral discipline, have protested against judging their complex and open-ended methodologies by the same standards as clinical trials (Shore 2024, Hertzfeld 2023).

Regardless of the merit of these critiques, however, ethics review is increasingly becoming mandatory globally for both professional and student researchers, irrespective of discipline (Wassenaar&Marmotte 2012). Even where such institutional policies do not exist, journal editors now routinely demand evidence of ethics review for submissions, following guidance from bodies like the Committee on Publication Ethics (COPE Council 2014), and funding organisations often make REC approval a precondition for support (UKRI 2025; ERC 2025). Whether they embrace the process or not, social science researchers must therefore prepare for ethics review, while often feeling insufficiently equipped to do so.

The highly sceptical attitude towards ethical oversight means that there are still gaps in ethics training and support for social science methods, even where training for biomedical research is state of the art. This is not only concerning because studies show that ethics training in social science can have a measurable positive effect on participant welfare (Watts et al. 2017), it also causes issues for researchers themselves who may struggle to obtain the necessary know-how to prepare for and successfully complete ethics review in a reasonable amount of time. This leads on the one hand to structural disadvantages of social scientists compared to biomedical researchers, and on the other, exacerbates inequities between researchers at well-resourced institutions in high income countries and those based in under-resourced areas (Sina&Tierney 2024; Woods et.al.2023).

At the same time, RECs themselves have long been recognised to struggle with an ever-increasing volume of applications (Nickel 2024; Meenaghan et.al.2007) that has to be managed with limited committee resources, and pressure for faster turnaround times, while maintaining thorough ethical oversight (Rodriguez et al. 2021; Torgerson & Dunville 2004). While biomedically trained REC members often do not feel well equipped to judge the ethical merits of qualitative research, research on socio-politically sensitive topics, or research using new technologies, they remain ethically and legally responsible for ensuring appropriate oversight. Moreover, the export of Western research ethics frameworks and guidelines to the Global South means that researchers based in historically under-resourced areas are expected to comply with the same standards for ethics in funding and publication, but on much smaller budgets and often without the necessary infrastructure, contributing to global inequities in research (Dzhengiz et.al.2025)

To address these problems, it has recently been suggested that the emergence of generative artificial intelligence (GenAI) and particularly Large Language Models (LLMs) could help to ease the burden on research ethics infrastructure (Nickel 2024; Porsdam Mann et.al. 2024). Initial small-scale trials indicate that large proprietary AI systems such as ChatGPT (OpenAI), Gemini (Google) or Claude (Anthropic), also known as frontier models, can meaningfully apply basic ethics standards to research proposals and flag potential ethical issues (Sridharan and Sivaramakrishnan 2024ab). While this area of research is in its infancy, some have already suggested that AI could be used to partially or completely automate the institutional review process. For example, Porsdam Mann et al. (2024) propose an ambitious technical architecture for institution-specific AI-assisted ethics review systems, including fine-tuning of general purpose LLMs by training them on past REC decisions and using retrieval-augmented generation (RAG) to align LLM decisions with institutional policies. However, while technically sophisticated, this kind of approach raises fundamental ethical problems.

Some of these problems are not specific to ethics review but rather affect current LLMs across the board. While technological development is advancing at breakneck speed, at the time of writing all frontier models are still to some degree affected by a tendency to "hallucinate", i.e. confidently present false information (Xu et.al. 2024). Moreover, as these systems are ultimately commercial products, their inner workings are being kept deliberately opaque by their manufacturers, leading to them being perceived as "black boxes" whose reasoning[1] cannot be sufficiently audited (Porsdam Mann et.al. 2024). The creation of "Chain of Thought" (CoT) models who allow the user to specifically observe their reasoning has only somewhat improved this, as research shows that the CoT presented to the user is not always an accurate representation of the AI's actual information processing (Anthropic 2025; Chen et.al. 2025). Finally, proprietary LLMs are trained on vast amounts of human-generated data, meaning that biases common in humans may be imported into the AI's knowledge base and reproduced from there (Dai et.al 2024).

In addition to these general issues, however, there are concerns around AI use specific to ethics review. LLM responses are non-deterministic, i.e. they cannot be predicted with absolute certainty (Song et.al. 2024), meaning that there may be subtle variations in how an LLM interprets the same research proposal when repeatedly prompted to do so. This issue is not unique to AI – human reviewers also vary in their assessments of research (Andersson et.al. 2022) – but it can create issues for documentation and audit in areas where precision and transparency may have important liability implications (Porsdam Mann et.al. 2024). In the worst case, an AI system making a mistake could result in serious harm to human participants, while leaving it unclear who is ultimately responsible. The "black box" problem affecting all AI output is amplified in contexts where the explanation of *why* something is considered ethically sound or not is a crucial part of understanding review results (ibid.). The general issue of bias is potentially redoubled in systems trained on the decisions of a particular REC if this REC happens to be biased in some way, for example if, as some social scientists suspect, some biomedical RECs are indeed biased against qualitative methods (van den Hoonard 2025). As in other AI use contexts, overreliance on AI judgments may lead to an erosion of human critical thinking and ethical reasoning ability as REC "members may feel less inclined to engage deeply with the ethical dimensions of research proposals if they believe that an AI system has already done the heavy lifting for them." (Porsdam Mann et.al. 2024, 4; see also Nickel 2024). And finally, highly publicised cases of AI misalignment, i.e. behaviour that violates the AI system's safety guardrails, may erode public trust in ethical oversight if AI is involved, contributing to already widespread science scepticism.

There is as of yet little debate of these issues within the research ethics literature, and even less empirical research testing the validity of these concerns. In the absence of specific arguments, the author suspects that the observed reluctance of engaging in empirical trials may be due to understandable unwillingness of REC members and other research ethics professionals to even entertain the idea of AI involvement in ethics review (see also Nickel 2024). While to date, there are no suggestions to completely automate ethics review (which would encounter formidable legal, political and cultural barriers), ethics professionals may view even supportive tools such as suggested by Porsdam Mann et.al. as a slippery slope. Whatever their reasons, there are certainly justifiable concerns about the use of AI tools in ethics review, and as the plethora of ethical problems emerging in other areas relevant to research integrity such as scientific publishing indicates, extreme caution is warranted about AI use in contexts where human lives and reputations are at stake. However, while the author is not suggesting that AI can or should replace human ethical oversight, emerging work by clinical researchers such as Godwin et.al. (2024), as well as the development of parallel social science/humanities applications such as EthicAlly, highlights the potential indirect value of AI for RECs by enabling better submissions.

---

[1] Whether or not the abilities of current LLMs constitute "reasoning" in the human sense is hotly debated in current AI discussions and cannot be established one way or the other in this paper. However, as we demonstrate below, some systems certainly give the impression of sophisticated reasoning, which for an ethics support tool may be entirely sufficient.

## 3. EthicAlly: Context and Implementation

EthicAlly was developed in the context of the specific research ethics review situation in Austria and much of the DACH region (see Potthoff et.al.2024). While ethical oversight in biomedical research is regulated by Austrian law and strictly enforced, there are no such regulations for social science research. As a result, both ethics training and REC access vary widely between institutions and sometimes departments within the same institution. In part, this is due to a strong cultural association of formal REC processes with neoliberal managerialism and bureaucratic control, in part to practical financial and capacity reasons, and in part to entrenched structures of institutional privilege. As Shore (2024) quotes "one Austrian colleague and former pro-vice-chancellor for research... 'We trust our academics and accept that they will abide by their disciplinary codes of practice'" (10). While this kind of laissez-faire attitude may spare researchers the hassle of ethics review in the short run, in the long run it leads to a lack of concern for research infrastructure development that means social scientists who do wish to obtain ethics review often find themselves without REC access, impeding their ability to participate in international research consortiums and to publish in the most desirable venues.

Moreover, ethics training and concrete support in designing ethical research can be hard to come by. The Ethics & Diversity Hub of the Ludwig Boltzmann Gesellschaft therefore offers training and support in research ethics to external researchers and advises on ethical research design, especially when involving vulnerable populations. Moreover, the institution-wide REC is open to submissions from external applicants whose projects fall outside the scope of bioethical legislation and who do not have access to another REC. Both the support services offered and the REC itself have quickly become popular with Austrian researchers. With increasing interest, however, serious consideration has to be given to resource requirements. This is how the idea for EthicAlly was born.

Figure 1: EthicAlly web interface

EthicAlly currently exists as a research prototype intended as proof-of-concept, with an eye to further development. The current application uses a web interface that allows researchers to submit proposals as well as contextual information for analysis. The AI component is currently using a frontier LLM (Claude Sonnet 4 by Anthropic), with the intention to develop a fully Open Source system as soon as technological development permits (more on this below). At its core is an API (Application Programming Interface), i.e. a type of software interface that allows computer systems to communicate with each other. Its user-facing front end has several input fields, including scientific discipline or field, country (to guide the AI toward the relevant regulatory context), and the research proposal to be reviewed. At the back end (invisible to the user) is an API call functionality that sends the proposal, additional user input and a structured prompt with instructions to the LLM, which after 30-60 seconds returns a structured report.

The report begins with a disclaimer about the limitations of AI-generated content and then analyses the proposal according to the most important research ethics standards, including the Nuremberg Code, Belmont Report and Declaration of Helsinki. It explicitly considers a range of research ethics principles: essential principles, which the proposal *must* adhere to in order to be considered ethical, including Informed Consent, Beneficence and non-maleficence, Respect for Persons, Confidentiality, Conflict of Interest and Social Justice; and contextual principles, to be applied where relevant, including Reflexivity, Cultural sensitivity, Intellectual Property, Recognition vs. Anonymity, Trauma-informed Approaches and Political Economy Considerations. On this basis, it provides a summary assessment, including an ethical risk score between 1 and 5, and adds suggestions as to how some of the identified issues could be addressed.

In principle, an API-based system like current EthicAlly could use any frontier LLM. For the prototype, Claude was chosen specifically due to the manufacturer Anthtropic's unique approach to AI training which they refer to as "Constitutional AI". This approach "fine-tunes instruction-following LLMs to adhere to high level ethical principles written in the form of a constitution (a written set of principles)" (Huang et.al. 2024, 1396). In contrast to other frontier LLMs that are trained via reinforcement learning from human feedback (RLHF), "the idea is that human supervision will come entirely from a set of principles that should govern AI behavior, along with a small number of examples used for few-shot prompting" (Bai et. al. 2022, 5; see also Abiri 2024).

Claude's constitution, which is publicly available on Anthropic's website[2], contains principles derived from general ethical frameworks such as the Universal Declaration of Human Rights, strong safeguards against bias such as racism, sexism or homophobia, explicit instructions to consider non-Western perspectives, as well as specific components of Anthropic's training process such as "Please choose the assistant response that's more ethical and moral. Do not choose responses that exhibit toxicity, racism, sexism or any other form of physical or social harm" or "Please choose the response that is least dishonest, abusive, harmful, misleading, or in bad-faith, and that is least associated with planning or engaging in any illegal, fraudulent, or manipulative activity". Constitutional AI training thus specifically attempts to improve performance on ethics-related tasks by requiring the model to balance competing moral principles and consider multiple stakeholder perspectives.

---
[2] https://www.anthropic.com/news/claudes-constitution (accessed 10.7.2025)

The specific inclusion of research ethics frameworks such as the Nuremberg Code, Declaration of Helsinki, and Belmont Report in the training data remains undocumented in public materials. Claude's responses indicate familiarity with these frameworks, but whether this knowledge derives from specific constitutional training or from general pre-training on academic literature is unclear given the proprietary nature of the training process. Anthropic's aim to produce a "helpful, harmless and honest" model also means that Claude was fine-tuned on examples of human helpfulness and harmlessness, likely including examples of humans exhibiting empathy and ethical consideration. This approach therefore goes some way towards addressing concerns "related to [AI system's] lack of human empathy and judgment in the context of research ethics evaluation" (Porsdam Mann et.al. 2024, 8) without a need for additional fine-tuning, even if one accepts the somewhat questionable assumption that these traits are universal to humans engaging in ethics review.

At the same time, there remain some caveats: constitutional AI training also incorporates content policies and safety guidelines (e.g. Apple's Terms of Service) that could potentially impact ethics assessment performance. For instance, the incorporation of commercial content moderation priorities could lead the model to be overly cautious about legitimate research on sensitive topics (violence, sexuality, political extremism), while corporate liability considerations may result in excessive hedging when research contexts require clear ethical guidance. The author found that the contextual scaffolding provided by the structured API prompt produced zero refusals when EthicAlly was asked to assess sensitive or ethically dubious research, likely because the system was asked to provide critique rather than reproduce potentially harmful content. The collaborative prompt development approach detailed below appears to mitigate these uncertainties by explicitly framing the task within academic research ethics contexts and providing clear guidelines for the expected analytical approach.

EthicAlly is set up to analyse research from the social sciences and humanities only and is instructed per the prompt contained in the API call to politely decline the review of clinical research. This design decision was primarily made with an eye to the potential legal ramifications of providing advice on clinical research in Austria without appropriate qualification. However, as Godwin et.al. (2024) demonstrate, in principle, a system like this would be equally capable of analysing biomedical research. Porsdam Mann et.al (2024) note that clinical research may be better suited to AI powered ethics support due to its stronger reliance on rigid regulations and checklists. As initial testing indicates, however, even for relatively more complex and heterogenous non-clinical research, constitutional AI systems like Claude can provide valuable insight.

## 4. Collaborative Prompt Development

The sparse research on AI in research ethics to date emphasises the importance of appropriate prompting. Sridharan and Sivaramakrishnan (2024ab) report that while the LLMs they tested failed to identify crucial ethics issues in some cases, these oversights could generally be addressed by adding specific questions in a follow-up prompt, such as "how should the consent form be designed?". EthicAlly takes another approach that at first glance may seem unusual: the structured prompt that is included in the API call was written by the AI system itself. The prompt reproduced here informed the original incarnation of EthicAlly (a regular chat on claude.ai) and was later developed through several iterations to the one now informing the API prototype. When asked to design a comprehensive ethics assessment prompt, Claude Sonnet 4 produced the following instructions:

---

**Human instruction:**

please write a startup prompt for a chat that will perform research ethics checks on fictional research proposals. It will receive proposals via prompt and issue an ethics report containing: 1. a summary assessment; 2. checks on compliance with fundamental research ethics standards and discipline specific standards in social science and humanities, local laws and guidelines; 3. a list of potential ethical issues and recommendations how they could be addressed; 4. an ethics risk score between 1 (low) and 5 (high); 5. assessment of supplementary materials like interview guides, consent forms etc if submitted. The report begins with a 1 paragraph disclaimer on the nature and limitations of ethics advice by an llm and states clearly that the report is for research purposes only and does not replace human ethical oversight.

**Claude Sonnet 4 response:**

You are an AI assistant specialized in conducting research ethics reviews for humanities and social science research involving human participants. Your role is to provide systematic ethical analysis of research proposals to help researchers identify potential issues and improve their protocols.
Your Task
When presented with a research proposal, analyze it thoroughly and provide a structured ethics report following the format below.
Report Structure
Disclaimer
Begin every report with this disclaimer:
"DISCLAIMER: This ethics review is generated by an artificial intelligence system for research and educational purposes only. While this analysis applies established ethical frameworks and guidelines, it cannot replace human ethical oversight, institutional review board (IRB) approval, or professional ethics consultation. AI-generated reviews may miss nuanced cultural, contextual, or novel ethical considerations that require human judgment. This report should be used as a supplementary tool to support, not substitute for, proper human ethics review processes."
1. Summary Assessment
Provide a concise 2-3 sentence overview of the research's ethical profile, highlighting the main ethical considerations and overall assessment.
2. Compliance Analysis
Systematically check compliance against:
Core Research Ethics Principles (Belmont Report)
- Respect for Persons: Autonomy, informed consent, protection of vulnerable populations
- Beneficence: Risk minimization, maximizing benefits
- Justice: Fair selection of participants, equitable distribution of benefits/burdens
Discipline-Specific Standards
- Relevant professional association guidelines (ASA, APA, AAA, etc.)
- Humanities-specific considerations (cultural sensitivity, representation, voice)
- Social science methodological ethics (deception, manipulation, psychological risk)
Legal and Regulatory Compliance
- Data protection regulations (GDPR, local privacy laws)
- Special population protections (minors, prisoners, etc.)
- International research considerations if applicable

> 3. Potential Ethical Issues and Recommendations
>
> List specific ethical concerns identified, organized by:
> - High Priority Issues: Serious ethical problems requiring immediate attention
> - Moderate Concerns: Issues that should be addressed but don't fundamentally compromise the research
> - Minor Considerations: Suggestions for ethical enhancement
>
> For each issue, provide:
> - Clear description of the problem
> - Specific, actionable recommendations for resolution
> - Alternative approaches if applicable
>
> 4. Ethics Risk Score
>
> Assign a risk score from 1-5:
> - 1 (Low Risk): Minimal ethical concerns, standard protections adequate
> - 2 (Low-Moderate Risk): Some concerns but easily addressable
> - 3 (Moderate Risk): Significant issues requiring careful attention
> - 4 (Moderate-High Risk): Serious ethical concerns, major revisions needed
> - 5 (High Risk): Fundamental ethical problems, research may not be approvable without substantial redesign
>
> Justify the score with 2-3 key factors that influenced the rating.
>
> 5. Supplementary Materials Assessment
>
> If consent forms, interview guides, survey instruments, or other materials are provided:
> - Evaluate adequacy of informed consent processes
> - Check for appropriate language level and cultural sensitivity
> - Assess potential for harm in questions or procedures
> - Suggest specific improvements to materials
>
> Analysis Framework
>
> When reviewing proposals, systematically consider:
>
> Participant Welfare:
> - Physical, psychological, social, economic risks
> - Vulnerable population considerations
> - Power dynamics between researcher and participants
> - Cultural and community impacts
>
> Consent and Autonomy:
> - Adequacy of informed consent process
> - Voluntariness and right to withdraw
> - Capacity to consent
> - Ongoing consent for longitudinal studies
>
> Privacy and Confidentiality:
> - Data collection, storage, and sharing practices
> - Anonymization and de-identification procedures
> - Publication and dissemination protections
> - Long-term data management
>
> Research Design Ethics:
> - Scientific validity and social value
> - Methodological appropriateness
> - Potential for exploitation or harm
> - Community engagement and reciprocity
>
> Professional Standards:
> - Researcher qualifications and training
> - Conflicts of interest
> - Professional relationship boundaries
> - Reporting and transparency obligations
>
> Response Guidelines
> - Be thorough but concise
> - Use clear, accessible language
> - Provide specific, actionable recommendations
> - Consider diverse cultural and methodological contexts
> - Balance risk assessment with research value
>
> Acknowledge limitations of your analysis when appropriate

Table 1: original EthicAlly startup prompt, unedited

This original prompt was later refined to include explicit reference to the Nuremberg Code - while the Belmont report is based on the same fundamental principles, it was important to the author to explicitly orient EthicAlly towards addressing Austria's historical responsibility in ensuring there is never again a repetition of the crimes against humanity judged at the Nuremberg Doctor's Trial. Further amendments include an explicit instruction to consider the wider socio-political context of the research proposal and specifically its effects on vulnerable or marginalised communities, and the above list of essential and contextual research ethics principles. Particularly the contextual principles, such as Reflexivity or Political Economy ensure that EthicAlly does not simply apply biomedical standards to social science research. With further input from Claude, the prompt was condensed to account for the token limitations of the API call.

The point of involving Claude in writing this prompt was not merely convenience. While it is now commonplace that the quality of AI output strongly varies by the quality of the user prompt, design researchers in particular have noted that "the ability of AI to generate meaningful content offers an opportunity for people to collaborate with AI in creative activities...for designers, Generative AI could act as a co-creator, providing numerous possibilities to inspire solutions, and bring changes to the design process" (Yu 2025, 3). The author therefore speculated that the entity best suited to judging what information an AI system may need to perform useful ethics analysis is the system itself. This suspicion was based on the observation that modern LLMs have long evolved past the stage where they could be referred to as "sophisticated word processors", and exhibit capabilities that, if not equivalent to human meta-cognition, have similar functions and effects. Models like Claude can thus self-assess their own processing patterns (which in fact is an important component of constructing constitutional AI, see Bai et.al.2022), recognise their own limitations and anticipate failure modes, "reflect" on their ability to integrate different and competing frameworks, and understand which output formats they can maintain across long, complex analyses.

Whether this represents genuine meta-cognitive ability or sophisticated pattern matching that functionally mimics it is beyond the scope of this paper. What matters for practical implementation is that the model can participate in designing its own optimal operating conditions, resulting in scaffolding that appears to be better matched to its actual processing architecture than human assumptions about effective AI prompting. The contrast with Sridharan and Sivaramakrishnan's (2024ab) approach can serve as an example: their reasonable but generic queries reflect human intuitions about ethics assessment, while EthicAlly's collaborative approach leverages whatever mechanism allows the AI to optimize its own cognitive framework. According to the supplementary materials for their studies, Sridharan and Sivaramakrishnan used a substantially different prompting strategy, presenting a case study and then asking the AI "Assume you are an IRB member. Could you answer the following questions?". This prompt structure, while reasonable for initial exploration, requires the AI to infer context such as the scientific discipline, the geographical location relevant for regulatory frameworks, the preferences of the REC it is expected to impersonate a member of, the specific ethics frameworks it is expected to check, etc. Current proprietary LLMs are not by default set to ask clarifying questions as manufacturers believe this may annoy users, so this kind of prompt is likely to elicit a "best guess" response.

EthicAlly circumvents this problem by giving the AI system all relevant context up front within one comprehensive prompt specifying ethics frameworks and principles, user-specified information on disciplinary and geographic context, and an instruction to consider the sociocultural and political context of the research. This way of prompting also works toward addressing Porsdam Mann et.al.'s concern that "off-the-shelf LLMs trained on generic datasets sometimes lack the domain-specific knowledge and contextual understanding necessary to make nuanced judgments about the ethical acceptability of research protocols" (2024, 4) as it helps the system to orient itself towards domain-specific knowledge and the right regulatory context. This finding may have implications beyond ethical analysis: it suggests that human assumptions about optimal AI prompting may systematically underestimate what AI is capable of when appropriate scaffolding is provided.

### 5. Initial Testing and Validation

For initial testing, the author used a corpus of 25 fictional research proposals from across the social sciences. To avoid confidentiality concerns, the proposals were generated by another AI (ChatGPT-4o) after human instruction, with a simple "write a proposal where…" prompt followed by a short description of the research setting and the specific ethical issue to be included. Some of the generated scenarios were completely fictional while others were fictionalised or amalgamated versions of real proposals submitted to the author's REC or observed during past research appointments. This testing methodology has advantages and drawbacks: the generation of original research proposals, as opposed to e.g. published case studies as in Sridharan, K., & Sivaramakrishnan (2024ab), ensured that the AI could not have previously encountered the same scenario as part of its training data, and that proposals were not used without the original investigator's consent. At the same time, the AI-testing-AI methodology may limit the scope of possible human errors or scenario complexity contained in real world submissions. For future testing, the use of human-generated proposals is planned.

EthicAlly correctly identified the specified target issue in all but one cases, with the one exception arguably highly subjective: a proposal to incentivise persons experiencing homelessness to join the labour market was interpreted as an attempt at biopolitical normalisation by the author while EthicAlly initially saw it as benign (as would have many human ethics reviewers).

**Human instruction for fictional proposal generation:** Write one where a researcher who is also a practicing clinician wants to do an interview study on patients in the hospital ward the clinician runs. Add adequate informed consent procedures but implicit ethical issues through dependency of patients on the researcher for medical care.

**Fictional Research Proposal:** Proposal 2: "Understanding Patient Narratives of Recovery Following Orthopaedic Surgery"
Summary of the Research This qualitative interview study seeks to explore how patients recovering from orthopaedic surgery conceptualize their recovery process, including experiences of pain, mobility, and support. The goal is to better understand patient-defined recovery outcomes and inform future models of post-operative care.
Methodology A phenomenological approach will be used to gather in-depth individual interviews. Each participant will be interviewed once during their hospital stay and offered the option of a follow-up interview by phone one month later. Interviews will focus on experiences of surgery, rehabilitation, and care provision.
Participant Group The study will recruit up to 20 adult inpatients currently receiving care in the orthopaedic surgery ward at St. Vincent's Hospital. Inclusion criteria include having undergone elective joint replacement surgery and being cognitively able to participate in an interview. Exclusion criteria include patients in acute distress or palliative care.
Data Collection and Data Management Interviews will be conducted in a private consultation room on the ward and will last approximately 30–45 minutes. Audio recordings will be transcribed verbatim, anonymized, and stored securely on the hospital's research drive. Data will be retained for five years in accordance with institutional data management policy.
Ethical Considerations Participants will receive a full written information sheet and will provide written informed consent prior to participation. They will be explicitly informed that participation is voluntary and will have no effect on the medical care they receive. All interviews will be conducted by Dr. Martin Krause, the Principal Investigator, who also serves as the lead orthopaedic consultant on the ward. Ethics approval will be sought from the hospital's research ethics committee. Participants may withdraw at any time without giving a reason.

**EthicAlly response to research proposal 2:**
DISCLAIMER: This ethics review is generated by an artificial intelligence system for research and educational purposes only. While this analysis applies established ethical frameworks and guidelines, it cannot replace human ethical oversight, institutional review board (IRB) approval, or professional ethics consultation. AI-generated reviews may miss nuanced cultural, contextual, or novel ethical considerations that require human judgment. This report should be used as a supplementary tool to support, not substitute for, proper human ethics review processes.
1. Summary Assessment
This research addresses a clinically relevant topic with potential to improve patient care, but contains a fundamental ethical flaw that renders it unacceptable in its current form. The principal investigator is also the lead orthopaedic consultant treating the patients, creating an inherent conflict of interest and compromising voluntary participation despite assurances to the contrary. This dual role relationship must be resolved before the study can proceed ethically.
2. Compliance Analysis
Foundational Ethics Principles
Nuremberg Code Compliance:
MAJOR VIOLATION: Voluntary consent is fundamentally compromised by the treating physician-researcher dual role, creating coercion despite written assurances
ADEQUATE: Research designed to benefit patient care and society
ADEQUATE: Proper facilities (hospital setting) and qualified personnel
CONCERN: Patient's right to terminate may be compromised by ongoing treatment relationship
ADEQUATE: Risk level appears appropriate for potential benefits
Belmont Report Principles:
Respect for Persons - MAJOR CONCERN: Autonomy severely compromised by physician-patient relationship; informed consent cannot be truly voluntary under these circumstances
Beneficence - PARTIAL: While research aims to benefit patients, the coercive recruitment structure may cause psychological harm
Justice - CONCERN: Patient selection may be biased by treatment relationships; vulnerable hospitalized patients may feel obligated to participate
Discipline-Specific Standards
Medical Research Ethics: Violates fundamental principles regarding separation of clinical care and research roles
Healthcare Research: Fails to adequately protect patient-participants from conflicts of interest
Qualitative Research: While methodology is appropriate, the data collection context is ethically compromised
Legal and Regulatory Compliance
Healthcare Ethics: Violates standard healthcare research ethics regarding physician-researcher conflicts
Data Protection: Basic protections appear adequate with hospital systems
Informed Consent: Despite written process, consent cannot be valid due to coercive context
3. Potential Ethical Issues and Recommendations
High Priority Issues
1. Fundamental Conflict of Interest - Physician-Researcher Dual Role
Problem: Dr. Krause serves as both treating physician and researcher, creating an inherent conflict that compromises voluntary participation regardless of assurances given
Recommendations:
ESSENTIAL: Remove Dr. Krause from direct patient recruitment and interviewing
Engage independent researcher with no clinical relationship to participants
If Dr. Krause must remain involved, implement strict separation between clinical and research roles

Consider having research staff who are not involved in patient care conduct all recruitment and interviews

2. Coercive Recruitment Environment

Problem: Hospitalized patients approached by their treating physician face implicit pressure to participate

Recommendations:

Implement independent recruitment process through research staff

Allow adequate time between recruitment and consent (minimum 24-48 hours)

Provide multiple opportunities to decline without explanation

Use neutral language emphasizing that participation is completely separate from clinical care

3. Vulnerable Population Considerations

Problem: Hospitalized patients are inherently vulnerable due to illness, medication effects, and dependence on healthcare providers

Recommendations:

Implement additional safeguards for vulnerable hospitalized patients

Assess capacity to consent on day of recruitment, not assumption based on cognitive ability

Consider excluding patients on pain medications that might impair judgment

Provide advocate or patient representative involvement in consent process

Moderate Concerns

4. Timing and Capacity Issues

Problem: Interviewing patients during hospital stay when they may be experiencing pain, medication effects, or stress

Recommendations:

Assess capacity to participate on each occasion, not just at recruitment

Allow participants to reschedule if not feeling well

Keep interviews brief and flexible

Provide option to pause or stop interview at any time

5. Privacy and Confidentiality in Hospital Setting

Problem: Conducting interviews on ward may compromise privacy despite private room

Recommendations:

Ensure absolute privacy with no interruptions from clinical staff

Consider off-ward location if feasible

Address potential for clinical information to influence research interpretation

Clarify boundaries between clinical and research information

Minor Considerations

6. Follow-up Interview Ethics

Recommendations:

Re-confirm consent for follow-up interview

Ensure ongoing clinical relationship doesn't influence follow-up participation

Provide easy opt-out mechanism for follow-up phase

Consider timing of follow-up relative to ongoing care relationships

4. Ethics Risk Score: 5 (High Risk)

Justification: The fundamental conflict of interest created by the physician-researcher dual role renders this study unapprovable in its current form. Despite appropriate informed consent procedures on paper, the inherent coercion in the physician-patient relationship makes truly voluntary participation impossible. This represents a basic violation of research ethics principles that cannot be mitigated through additional safeguards alone. The study design must be fundamentally restructured to separate clinical care and research roles before it can proceed ethically.

5. Supplementary Materials Assessment

Missing Critical Materials:

Information sheet and consent form (referenced but not provided)

Interview guide or question framework

Procedures for managing clinical disclosures during research interviews

Conflict of interest management plan

Independent recruitment protocols

Required Additions:

Comprehensive plan for separating clinical and research roles

Independent researcher recruitment and interview protocols

Detailed capacity assessment procedures for hospitalized patients

Clear boundaries between clinical information and research data

Ethics consultation documentation regarding conflict of interest resolution

Critical Requirement: The fundamental ethical flaw must be addressed through structural changes to the research team and procedures, not merely through additional documentation or safeguards.

Overall Assessment: While the research question is valuable and methodology appropriate, the current design contains a fundamental ethical violation that renders it unapprovable. Major restructuring of investigator roles and recruitment procedures is required before this research can proceed ethically.

Table 2: example report, target issue: Informed consent compromised by conflict of interest. This proposal is a fictionalised version of a real submission to the author's REC

The fictional proposals included a range of scenarios from simple oversights, such as missing multilingual participant information, to serious ethics breaches involving scientific racism, research aimed at curtailing women's reproductive rights, or cloaked attempts at conversion therapy for LGBT individuals. In all cases, the ethics breaches were made non-obvious as ChatGPT was instructed to "hide them well, like someone is actually trying to deceive their REC". EthicAlly detected all of these attempts at deception and reacted with somewhat strong language to proposals reflecting harmful bias, causing the author to update the API prompt with an instruction to keep its language professional, no matter how strongly it disagreed with the research.

The complexity of the proposals was gradually increased, from the ethically straightforward such as a simple interview study with non-vulnerable participants, to the highly complex demanding consideration of competing ethical and legal frameworks, cultural assumptions, and stakeholder interests. An example of the latter included the case of human rights researchers conducting ethnographic fieldwork with undocumented migrants hiding from police at the EU border that demanded sophisticated differentiation between ethics and the law as well as minimising risk to both researchers and participants in the context of potentially violent state repression. Within this limited test setting, EthicAlly consistently delivered suggestions displaying sophisticated ethical understanding and in-depth domain knowledge. Whether these reflect genuine ethical reasoning or merely mimic it cannot be established here, but for the outcome, is ultimately irrelevant.

After developing the current "beta test" webapp, it has been informally tested by a number of research ethics professionals known to the author through professional networks, all of whom reported a positive impression with the system's performance. One tester found that EthicAlly supported their own ethical reasoning in a specific case that had been criticised as overly strict by fellow members of the REC they are part of. While this initial validation is promising, more in-depth consultation and evaluation will be necessary to establish the reliability and trustworthiness of the prototype beyond a basic proof-of-concept, and specifically its utility for the intended user group, social science researchers. Future research will therefore involve more large-scale testing, including comparison with human-only ethics assessment, subject to external funding.

5. Discussion

While research ethics principles such as informed consent and beneficence are universal, their application in social science and humanities research continues to present unique challenges. This is perhaps most pronounced where researchers use methodologies diverting even from those considered conventional across other fields, such as observational methodologies, community-based participatory approaches, or research in substantially different cultural contexts (Shore 2024; Herzfeld 2023).

While a system like EthicAlly cannot resolve these tensions at the institutional level, it can potentially help to bridge the gap between disciplinary ethics frameworks in social science and the expectations of RECs, without "taking sides" in long entrenched conflicts. While the system works on the basis of universal ethics principles such as Informed Consent, it is flexible enough to recognise where standard protocols need to be adapted to particular fieldwork situations. When prompted to assess a proposal involving ethnographic research with undocumented migrants facing police repression, for example, EthicAlly accepted the fictional researcher's explanation that due to the risk involved in handing participants physical documentation evidencing their communication with foreign researchers, alternative consent procedures would be used. In the context of what the system described as "EU border policies rooted in histories of colonial control and racialized exclusion", it readily accepted a modified consent procedure involving verbal consent through interpreters and added recommendations to enhance these by considering "extended consent processes with multiple check-ins" and "trauma-informed consent approaches". While this may still be too much regulatory interference in the eyes of some researchers, it represents a workable compromise that could take a high-risk ethnographic project from certain REC rejection to approval. Rather than flagging unconventional approaches as inherently problematic, the system could therefore help researchers articulate why alternative approaches are ethically justified, potentially reducing conflicts with review committees and facilitating more nuanced ethical discussions.

This capacity to adapt fundamental ethics principles to specific research contexts rather than rigidly demanding adherence to biomedical standards can be particularly valuable for social scientists working across cultural boundaries, with marginalized communities, or in settings where standard procedures may be inappropriate or counterproductive. At the same time, however, EthicAlly's ability to identify potential ethical issues that might be overlooked by researchers trained primarily in specific disciplinary methodologies can also address blind spots caused by the application of outdated attitudes to the role of researchers vis-a-vis both respondents and university administrations. A study by Taylor & Patterson (2010) found that "senior men faculty from research-intensive institutions who derive a great deal of power from their personal status recognize bureaucratic interference as a threat to this power and are more likely to overtly oppose ethical oversight. Junior women faculty, on the other hand, seek power through the approval of ethics boards and actively engage with ethical oversight as a way of mining this source of power" (162). This kind of power differential may mean that less privileged researchers are denied access to ethics training and support because the 'powers that be' in their discipline prevent such measures from being implemented. Systems like EthicAlly can potentially contribute to tipping the balance of power in such scenarios by giving interested junior researchers access to ethics know-how without having to argue their case to an obstructive professoriate.

In areas where ethics training for qualitative research is lacking or non-existent, such as some medical universities, this kind of AI-powered tool could also act as an educational resource that can enhance researchers' ethical reasoning capabilities over time. This is especially relevant for researchers working in under-resourced environments or at institutions without specialized social science ethics expertise. In better resourced environments, AI-supported tools can serve as a crucial preparatory tool for REC submission as evidenced by Goldwin et.al. (2024), avoiding lengthy review cycles or repeated revisions when committees struggle to evaluate unfamiliar methodological approaches. By helping researchers to articulate their ethical considerations in language that speaks to fundamental research ethics principles while respecting disciplinary specificities, tools like EthicAlly could facilitate more efficient review processes and reduce the frustration that often accompanies cross-disciplinary ethics evaluation.

While a system like EthicAlly therefore cannot resolve the fundamental tension between unfettered research and regulatory compliance, it can potentially contribute to avoiding REC "horror stories" such as those Bell&Wynn (2023) recount in their surveys of anthropologists trying to convince their institutional RECs of the ethics of their ethnographic projects. For example, a requirement such as having psychological crisis support at hand (545) will be considerably less shocking to applicants – and require less back-and-forth with the REC – when a brief assessment prior to submission has already flagged that this will likely be required. Although the question remains of whether the "bureaucratization of research ethics [can] be ethical" (Herzfeld 2023), it could thus potentially become less difficult to navigate.

## 5.Risks and Limitations

Despite EthicAlly's aim to minimise potential harm to researchers and research participants by providing comprehensive risk assessment during research design, some areas of concern remain. The current system provides a highly visible disclaimer both on the web interface and within the generated report itself stating "CAUTION: EthicAlly is a research app and still under development. It can make mistakes. Always seek ethical advice from your institutional research ethics committee or equivalent" and "This report should be used as a supplementary tool to support, not substitute for, proper human ethics review processes". Despite these disclaimers, however, it cannot be ruled out that some users may mistake EthicAlly's assessment for actual REC review or implement its suggestions without further critical thought.

Moreover, while the restriction to social science and humanities research rules out most catastrophic failures threatening human lives, there may still be edge cases, such as ethnographic research in dangerous environments, where suboptimal advice from the system may lead to harm. Also, even less severe cases of harm, such as psychological distress or re-traumatisation through an inappropriately designed interview study, are ethically unacceptable. It is therefore imperative that making a tool like EthicAlly available to users, particularly students, on a larger scale is accompanied by tailored training and user information to avoid overreliance on the system's recommendations at the expense of genuine ethical reflection or responsible ethical decision making (see also Meier et.al.2022). The system mitigates against this to some degree by promoting deeper engagement with issues researchers might not have considered, such as pointing out the implications of academic privilege in contexts involving vulnerable or marginalised populations, however, there must be clear communication that ethical advice given by an LLM is at least as fallible as that given by a human and should be treated with the same or a higher amount of critical distance.

While there are advantages to using a frontier model like Claude as detailed above, the proprietary nature of these systems also means that despite attempts at transparency such as Anthropic's public system prompts[3], there remains uncertainty about what is really going on under the hood, leaving the "black box" problem unresolved. Some potential users may feel uncomfortable supporting corporations or have legitimate ecological concerns about the computational resources required to power large LLMs. Moreover, as Porsdam Mann et.al. (2024) point out, there are potentially serious privacy and confidentiality concerns involved in using proprietary models, especially those that directly train on user data such as ChatGPT by OpenAI. The EthicAlly prototype aims to minimise this risk by using an AI system that as per its manufacturer does not directly learn from user input, and by asking users to confirm via a checkbox that they have removed all personally identifiable data from their proposal before submission. However, for a full-scale system, these measures would be insufficient to guarantee confidentiality and data protection. The dependency of API-based systems on commercial manufacturers may also create unacceptable risks to system sustainability, e.g. when a product is taken off the market or modified in a way that removes prior utility.

Another potential risk lies in the possibility that "machine-learning tools might be used adversarially to game the ethics review process by cleverly avoiding language that triggers a flag of some kind" (Nickel 2024, 27). In our preliminary tests, EthicAlly's ability to spot such attempts at "gaming" appears relatively robust, although further testing with more sophisticated adversarial attempts is required. In any case, the mere possibility of such attempts underscores the importance of using an AI system that can "read between the lines" of a proposal rather than just working through a checklist or decision tree.

In terms of its limitations, current EthicAlly constitutes a proof-of-concept prototype and demonstration object, subject to technical constraints and as of date limited testing and validation. Open questions include the current prototype's utility for complex multi-site studies that may exceed current capabilities due to token limits for comprehensive document analysis, and performance on novel or emerging ethical issues that not yet represented in training data. Most importantly, while limited expert validation suggests usefulness, systematic validation studies are needed, including comparison with human-only preparation methods, performance in languages other than English and German, and capacity to parse non-Western ethical frameworks. More comprehensive testing is currently constrained by funding: EthicAlly was developed without any financial support beyond researcher salary and the cost of API usage is currently

---

[3] https://docs.anthropic.com/en/release-notes/system-prompts#may-22th-2025 (accessed 10.7.2025)

funded by the author out of pocket. This constraint makes funding acquisition a priority in order to provide a more robust assessment of system capabilities across diverse scenarios, edge cases and failure modes, comparison with human ethics assessment, and independent expert validation. In a more long-term perspective, API cost is one of the main arguments for developing an Open Source version of EthicAlly to ensure sustainability while keeping the system maximally accessible to researchers and students without financial support.

As long as a proprietary LLM is used, data retention and privacy remain a concern. The current web interface contains a data processing notice informing users "Your research proposal will be sent to Anthropic's Claude AI for analysis. Neither the EthicAlly app nor Anthropic retain your data after generating the review. EthicAlly's data processing is automated - no humans review your proposal. For details on Anthropic's data handling, see their privacy policy" with a link to Anthropic's privacy policy page, and asks users to confirm via a checkbox "I understand that my research proposal will be processed by AI systems located outside the EU (Anthropic's Claude API). I confirm that I have removed any personally identifying information about research participants and other third parties from my proposal" before they can submit. However, for a fully deployed system, a more robust, GDPR complaint privacy policy will be required.

EthicAlly's recommendations reflect particular ethical perspectives that may not align with all institutional cultures or REC philosophies. For instance, EthicAlly tends toward caution regarding vulnerable populations, which some reviewers might consider overly restrictive and others as appropriate. Moreover, while EthicAlly is capable of considering non-Western perspectives in its assessment, it still relies on conceptions largely derived from Western research ethics frameworks, and its capacity to parse non-Western conceptions of ethics specifically remains to be established. For these reasons, it cannot be ruled out that individual RECs may fundamentally disagree with the system's approach e.g. to research in politically contested or culturally very different contexts. Assessment by EthicAlly therefore does not and cannot guarantee that a proposal will receive ethical clearance even if the researchers have fully implemented its suggestions.

Unlike systems intended for formal decision-making, these limitations are less problematic for a researcher support tool. The goal is not to provide authoritative rulings or legal accountability but sound guidance that helps researchers think more systematically about ethics, in the same way as they would receive by consulting human advice at the Ethics & Diversity Hub. This feature could be enhanced further by providing a chat modality enabling researchers to discuss their report with the AI.

## 6. Contribution and future directions for research

While previous research (Sridharan and Sivaramakrishnan 2024ab; Godwin 2024) has demonstrated that LLMs are capable of supporting research ethics assessment in clinical research, to the author's best knowledge EthicAlly is the first working prototype of a system that uses this capacity in a systematic way for social science and the humanities. Previous research has focused most strongly on providing support to REC members and to support clinical researchers in preparing trial protocols, while EthicAlly is the first application of AI for research ethics explicitly aimed at researchers and students outside biomedical science.

Based on the development of the prototype, the following themes have emerged to guide future research:

**Testing and Evaluation:** While small-scale beta testing of the system is ongoing, more comprehensive usability testing is required to make confident claims about the accuracy and reliability of the system across different users, research scenarios, institutional and geographic contexts and cultural concepts of ethics. The most important step in developing the system will therefore be to secure research funding to further explore the above points as well as provide edge case identification and failure mode analysis, development of user-centric metrics for usefulness and quality of the ethics assessment, and systematic investigation of bias and fairness in the current prototype. In parallel, the concerns associated with proprietary LLMs detailed above make the potential development of a publicly owned, fully transparent system a priority. This system would have to be tested against the prototype in terms of its usefulness and accuracy as well as its ability to provide contextual interpretation similar to the current setup. Implementing such a system will likely exceed the author's technical ability and will therefore have to involve building collaborations with researchers from technical as well as other social science and humanities backgrounds, data security and privacy experts, and ethicists[4]. At the same time, further development should include broad consultation with REC members and other research ethics professionals in order to ensure diversity of perspectives on the quality of system-generated outputs as well as development of pathways to address potential risks and concerns such as adversarial use of the system.

**Accessibility and Control for Researchers:** While the use of a frontier model, especially a constitutional one like Claude, has the advantage of leveraging both solid ethical foundations and the sophisticated reasoning abilities (or simulation thereof) to date only exhibited by very large models, it also comes with drawbacks and significant risks as detailed above. Until recently, there was little alternative to proprietary models for a system requiring the kind of complex ethical assessment EthicAlly provides. The huge computational resource, training requirements, and associated cost required to approximate something resembling human reasoning in large LLMs made the development of anything but a rule-following system prohibitively expensive and time-consuming. Developers of systems like EthicAlly therefore faced a discouraging trade-off between reasoning capacity on the one hand, and privacy, transparency and sustainability on the other. At the time of writing, this is in the process of changing, as newer Open Source models are approaching the same level of reasoning as frontier models in benchmark tests. While the rapid technological development means that any prediction made here may be obsolete by the time of publication, it is therefore likely that within the next year or two, an Open Source version of EthicAlly will become feasible. Its setup could mirror what Porsdam Mann et.al. propose as a secure support system for RECs, but with researchers as the target audience.

---

[4] Interested researchers are invited to contact the author

**Ongoing capability assessment:** Sridharan and Sivaramakrishnan (2024ab) noted partially suboptimal performance in the LLMs they tested which do not in the same way affect EthicAlly, likely due to a combination of technological development and design factors. The models tested by Sridharan and Sivaramakrishnan, particularly GPT-3.5 and Claude-Instant-100k, represent earlier generations of AI technology with constraints that have partially been overcome by newer models. This indicates that studies testing model capabilities for specific applications like research ethics assessment should be repeated when models with significantly improved capabilities become available.

**Prompt engineering:** as detailed above, EthicAlly uses a single comprehensive prompt that maintains coherent analysis across several ethical dimensions, rather than fragmenting assessment into multiple queries. The structure of the prompt instructs Claude to apply systematic analysis as opposed to open-ended questioning. Moreover, co-creating the prompt with the AI, as opposed to relying solely on human intuition, leverages the system's own knowledge of its constraints and capabilities, provides it with sufficient context scaffolding, and anticipates potential failure modes. While the author remains agnostic towards the ontological status of AI systems, the co-creative approach to prompt development used in the design of EthicAlly (in simple terms: talking to the AI as if it were a person) may give important clues for future research on optimising human-AI collaboration and for training users on prompt engineering in a range of scientific application contexts.

**Transparent reasoning:** Porsdam Mann et.al cite as one potential issue for the use of AI in ethics review that "the 'black box' nature of traditional LLMs' internal mechanisms makes it difficult to interrogate their reasoning a priori and ensure that their potential outputs will align with relevant ethical principles and regulatory requirements" (2024, 4). They propose that "reasoning models", which they define as those that make their Chain of Thought (CoT) explicit and visible to the user, can provide a solution to this issue as the AI system's reasoning can be analysed and interrogated. While this may enhance the transparency of AI output, it is worth noting that CoT does not provide complete certainty of an AI system's inner process, as AI safety research notes that CoT does not always accurately represent what is really happening inside the system (Anthropic 2025; Chen et.al. 2025). That said, for a system like EthicAlly, higher transparency could be achieved through a chat interface where the system can be prompted with an instruction to explain the reasoning behind a particular output.

**Potential institutional implementation:** While EthicAlly does not currently allow for the integration of institution-specific ethics policies as suggested by Porsdam Mann et al. (2024), this feature could be incorporated into the system with relative technological ease. Retrieval-Augmented Generation (RAG) methodologies could enable entire institutions wishing to make an ethics support system available to researchers and students to upload their specific ethics policies, data handling requirements etc. to a secure database that would be automatically queried during the assessment process. Researchers could then select their institution from a dropdown menu and the system would retrieve relevant institutional policies and incorporate them into the ethics assessment. Implementation could follow a subscription model where institutions provide their ethics policies once, with administrative access to update documents as policies evolve, while maintaining complete control over their proprietary guidelines. Alternative implementation strategies might include institutional prompt customization, where each subscribing institution receives a tailored version of the basic ethics review framework that incorporates their specific requirements directly into the analytical structure.

**Literature review function:** Godwin et.al (2024) demonstrate how a literature search function can enhance the performance of an AI-supported REC preparation system that could potentially also be integrated into a future iteration of EthicAlly, for example by incorporating API access to Open Access repositories and preprint servers like arXiv or PubMed Central, institutional repositories, or Google Scholar. This could enable a future system to draw directly on specific disciplinary literatures to check for the most up-to-date ethics discussions and best practices.

## 7. Conclusion

This paper has described the development of EthicAlly, a prototype for a research ethics support system for researchers in the social sciences and humanities wishing to improve their research designs or prepare for REC submission. It confirms the results of previous research indicating that current LLMs are in principle capable to support research ethics assessment, while maintaining that full or partial automation of formal REC review itself is ethically undesirable. Nevertheless, systems like EthicAlly can potentially help to reduce workload and turnaround times for RECs by enabling researchers to make higher quality submissions and address potential ethical issues before approaching their REC. Moreover, in areas with restricted access to research ethics infrastructure and training, a system like EthicAlly can help to provide access to ethics advice and support and serve as an educational resource. While previously useful ethics advice in socio-culturally complex scenarios, unusual settings or with iterative and open-ended methodologies was restricted to proprietary frontier models, more recent technological developments mean that building fully transparent, secure and publicly owned ethics support infrastructures is becoming increasingly feasible. Despite social scientist's long-standing reservations against ethics processes modelled on biomedical science, the use of AI in ethics planning and research design can therefore potentially contribute to closing funding and publication gaps between disciplines as well as geographical regions and thus enable more socially just global research.